\documentclass[runningheads]{llncs}
\usepackage[T1]{fontenc}
\usepackage{graphicx}
\usepackage{booktabs}
\usepackage{tabularx}
\usepackage{adjustbox}
\usepackage{float}
\usepackage{caption}
\usepackage{enumitem}

\begin{document}
\title{Evaluation of Large Language Models’ Educational Feedback in Higher Education: Potential, Limitations, and Implications for Educational Practice}
\titlerunning{Evaluation of Large Language Models’ Educational Feedback in Higher Education}
%
\maketitle Daniele Agostini1[0000-0002-9919-5391] and Federica Picasso1[0000-0002-8381-6456] 
\begin{abstract}
The importance of managing feedback practices in higher education has been widely recognised, as they play a crucial role in enhancing teaching, learning, and assessment processes. In today's educational landscape, feedback practices are increasingly influenced by technological advancements, particularly artificial intelligence (AI). Understanding the impact of AI on feedback generation is essential for identifying its potential benefits and establishing effective implementation strategies.
This study examines how AI-generated feedback supports student learning using a well-established analytical framework. Specifically, feedback produced by different Large Language Models (LLMs) was assessed in relation to student-designed projects within a training course on inclusive teaching and learning. The evaluation process involved providing seven LLMs with a structured rubric, developed by the university instructor, which defined specific criteria and performance levels. The LLMs were tasked with generating both quantitative assessments and qualitative feedback based on this rubric. The AI-generated feedback was then analysed using Hughes, Smith, and Creese's framework to evaluate its structure and effectiveness in fostering formative learning experiences.
Overall, these findings indicate that LLMs can generate well-structured feedback and hold great potential as a sustainable and meaningful feedback tool, provided they are guided by clear contextual information and a well-defined instructions that will be explored further in the conclusions.

\end{abstract}
\section{Introduction}
Feedback practices in Higher Education are largely identified as key elements to sustain in a tailored manner the learning process. It allows students to have a personalised support that have been demonstrated to be essential for improved performance and enhanced achievement on the task (Panadero and Lipnevich, 2022). Due to the importance of feedback for students’ academic and future professional growth, this study aims at exploring the emergent role which AI is playing in the field. Specifically, we see useful to consider whether, and to what extent, existing Large Language Models (LLMs) can be a valid support for teachers in the production of educational feedback.
Building on two previous studies (Agostini \& Picasso, 2024; Agostini, Picasso \& Ballardini, 2024), the research aims to analyse the features of the feedback produced by seven LLMs employed to assess the products of university students enrolled in a habilitation program for teaching in secondary school.
Through the use of the big-Agi platform (https://big-agi.com/) connected to all models’ APIs, we prompted simultaneously the seven LLMs asking generically, in a non-structured way, for feedback for each criterion on which the students’ product have been assessed. Asking feedback is such a non-structured, generic way, was fundamental to simulate the situation in which a teacher or an instructor could ask, without specific knowledge about feedback.
We chose to analyse the feedback produced through the use of Hughes, Smith and Creese's (2015) framework, that considers praise, critique, advice, and clarification as the four main features of a well-formed feedback.
The LLMs ability to provide well-formed feedback is then evaluated and compared thanks to a qualitative analysis and a coding procedure, to understand if LLMs can create powerful feedback, fit to scaffold students’ progress. If up to the job, these tools could represent a concrete help for university instructors in the management, and delivery of feedback processes. Being the feedback the single most impactful teaching tool in a teacher’s hand (Hattie and Yates, 2013) and being it often neglected due to time constraints and scalability issues, this could be the single most groundbreaking use of generative AI technology in education.

\section{Theoretical Framework}
\subsection{The Power of Feedback}
Many authors have provided a variety of definitions of feedback, but often associate it with the process by which learners acquire information about the quality of their own product, which results from comparing the product with an expected standard of quality. In this way, feedback enables improvement of the product or performance through change and improvement that will lead to future improvements and changes (Sadler, 1989). According to this perspective, the feedback that the teacher provides to students often offers corrective guidance and food for thought, enabling them to identify areas for future learning through a self-assessment of their current positioning and subsequent adjustment of their learning (Sadler, 1989; Nicol, 2010; Dochy, Segers and Sluijsmans, 1999) defined feedback as:

\begin{quote}
a process whereby learners obtain information about their work in order
to appreciate the similarities and differences between the appropriate standards for
any given work, and the qualities of the work itself, in order to generate improved
work (Boud and Molloy, 2013, p.6).
\end{quote}

Hattie and Timperley (2007) theorise four distinct levels of feedback classification that shed light on different aspects of how students engage with and respond to their learning. The first level, task-oriented feedback, directs attention to the specific performance of an activity by identifying what is correct or incorrect and by highlighting corrective actions for improvement. The second level, which addresses the process involved in completing the task, focuses on the interrelations among the various components that students mobilise, including the strategies they use to arrive at solutions. Moving to the third level, self-regulatory feedback fosters learners’ ability to detect and compare information, thereby prompting them to reassess and refine their approach. Finally, although many view positive remarks on the work or its author as inherently reinforcing, Hattie and Timperley (2007) contend that such generic praise tends to be counterproductive by diverting students’ attention from substantive factors, such as performance, processes, and self-regulation.

So, how can feedback be used to sustain students’ learning improvement?

Research has consistently shown that feedback is a key factor in student achievement and, if well-structured, can lead to improved performance on a task.
Hattie and Timperley (2007) argue that the most effective feedback bridges the gap between where a student currently is, the state of the goal they are aiming for, and the steps they need to take to get from the current state to the desired state. Shute (2008) also argues that feedback should help reduce uncertainty between performance and goal, as well as being supportive, timely, non-evaluative and specific (Lipnevich and Panadero, 2021).

Nicol and MacFarlane-Dick (2006) summarise the seven principles for crafting a good feedback:

\begin{enumerate}
    \item Feedback must help to clarify what good performance consists of in terms of objectives, criteria and expected standards; 
    \item It must facilitate the development of self-evaluation and reflection on learning for the recipient in training; 
    \item It must offer high-quality information to students regarding their learning; 
    \item Feedback encourages dialogue with the trainer and between peers and; 
    \item It encourages motivation and self-esteem; 
    \item It offers opportunities to reduce the gap between current and expected performance; and finally, 
    \item It offers information to teachers on how to adapt the learning path.
\end{enumerate}

Nicol sustains that well-formed feedback allows students to understand the meaning of the grade or judgement obtained in relation to the evaluation criteria used and shared. It allows them to recognise the strengths and weaknesses of their task/product and to identify future areas of learning, through a self-assessment of their current placement and a consequent adjustment of their learning (Nicol, 2010; Grion and Serbati, 2019, p.82).

Finally, for feedback to be truly effective and achieve a tangible impact on learning, several conditions must be met. Crucially, it should be delivered promptly so that students can act upon it while the task and its context remain fresh in their minds. Moreover, feedback needs to be frequent and sufficiently detailed to guide learners’ next steps. Equally important is the emphasis on learning objectives, rather than simply assigning marks, ensuring that feedback orients students towards progress and growth. In addition, linking feedback to specific learning outcomes and assessment criteria provides a clear framework for improvement, while ensuring that comments are comprehensible fosters deeper engagement with the material. Ultimately, feedback reaches its full potential when students actively receive and use it as a tool to refine their work and enhance their performance (Gibbs, Simpson and Macdonald, 2003; Grion and Serbati, 2019).

\subsection{AI and Assessment in Higher Education}
Nowadays, Large Language Models (LLM) have taken on a very significant role in the technological landscape. As a result, educational institutions and agencies have begun to incorporate LLMs and generative AI into their curricula at various levels, developing courses to harness the potential of these innovative technologies.
To address potential issues related to the use of AI in the educational field, national and international institutions, as well as universities, have established guidelines and frameworks promoting ethical use of LLMs. Organisations such as UNESCO (2023), the Joint Information Systems Committee (JISC) (2023), the Russell Group (2023), the French Ministry of Education (2023), the US Department of Education (2023),  the University College London UCL (2023) and the Agency for Digital Italy (AgID) (2025) have adopted this approach to structure specific guidance and introduce a scaffolding system for educational professionals and related institutions all around the world. 
Among the proposed AI applications into the design and the implementation of teaching and learning processes, assessment tasks appear to benefit most from AI, particularly in terms of sustainability. However, caution is essential, as LLMs without task-specific adaptations are unreliable for independently managing assessments. Importantly, instructors must consider ethical implications and act responsibly when evaluating AI-assisted tasks or if they adopt AI for assessment and feedback production, as these decisions can significantly influence students’ academic trajectories, including grades, motivation, and access to opportunities such as scholarships or advanced programs.
It is important to identify the advantages and difficulties associated with integrating AI systems into the evaluation procedures in educational settings by starting with the research of Kamalov et al. (2023). AI can assist academics solve time-consuming issues by introducing automation techniques that can drastically cut down on the time and effort needed for evaluation. By lowering the possibility of prejudice or human error, highlighting the significance of evaluation criteria, and encouraging correctness and fairness, automation can increase the dependability of assessment procedures (Bozkurt et al., 2024).
This is important for the authenticity process because AI appears to help teachers evaluate difficult tasks by automating feedback processes and assisting in the identification of certain addressing their learning gaps and simultaneously offering the opportunity for individualised instruction and evaluation (Kamalov et al., 2023; Minn, 2022; Kochmar et al., 2022).
On the other hand, there may be difficulties in implementing AI technologies.
Kamalov et al. (2023) draw attention to a potential problem with adaptability: automated assessment systems have trouble with original or imaginative student answers that deviate from preset answer patterns or rubrics.
During the assessment process, AI systems may find it challenging to recognise the sophisticated thinking of creative students. This could lead to misunderstandings or an incomplete evaluation, creating issues with contextual understanding of the assessment tasks and associated responses/products.
Although the capabilities of this technology are undoubtedly going to grow in the near future, it now has several shortcomings that need to be addressed (Picasso, Agostini, Serbati, 2025).

\section{The Current Study}
This study represents a natural progression of two previously developed studies (Agostini \& Picasso, 2024; Agostini, Picasso \& Ballardini, 2024) which aimed to examine the use of advanced Large Language Models (LLMs) for assessing student-written work, focusing on their accuracy and ability to evaluate based on teacher-designed grading rubrics. These studies' goal was to determine whether, and which, models can be effectively utilised by educators—both in university and non-university settings—who lack expertise in Machine Learning, to assess written products, including open-ended tasks, using grading rubrics.

This study, like the was conducted at the University of Trento as part of a university habilitation course for secondary school teachers, within a module on learning methodologies and distinguishes from the other two for its focus on determining the quality of the descriptive feedback given by LLMs.
Specifically, the research questions are the following:

\begin{itemize}
    \item[RQ1] Are LLMs able to produce descriptive feedback that supports students’ learning process effectively?
    \item[RQ2] Are LLMs feedback well-formed?
\end{itemize}

\section{Methodology}
\subsection{Participants}
The pilot study at the basis of the current research was developed with a sample which involved one-hundred-fourty-two students who took part anonymously in the training exercise: they were divided into 35 groups, along with 3 evaluating teachers, experts in experimental pedagogy and assessment. No
data regarding the students’ demographics was collected (Agostini, Picasso \& Ballardini, 2024).

\subsection{Task}
The task involved redesigning a past educational intervention that had failed, tailored to a specific class (ranging from the 1st grade of lower secondary school to the 5th grade of higher secondary school, depending on group composition). Groups were required to identify the original teaching approaches and strategies, propose alternatives better suited to achieving the intended learning outcomes, and reflect on their redesign process. 

\subsection{Instruments}

\subsubsection{For Assessing Students' Task}

\paragraph{Rubric}

The evaluation of the students’ task was based on a specific rubric (Appendices 1), and students were provided with a structured template for their educational design, which included sections on Involved Disciplines, Class and Grade Level, Intervention Title, Teacher, Programme and Learning Objectives, and Context and Environment (formal or informal, type of setting, etc.). The groups were given two hours and thirty minutes to complete the task.
Additionally, the groups were required to describe the reflection process they applied in redesigning the educational intervention, which was also evaluated. In the detailed section of the schedule, they provided concise explanations of the planned educational activities, specifying the roles of both teachers and students. Within this structure, the groups had the freedom to propose their original programming. Each group submitted a final product in the form of an MS Word document containing the educational intervention programming, structured according to the provided template.
For evaluation, a grading rubric with five criteria, each featuring four levels, was developed.

\paragraph{LLMs as Assessors}
The final products were assessed by three expert human evaluators and seven Large Language Models (LLMs), along with an additional model that combined the feedback of all seven LLMs into a single evaluation (GPT 4.0). The LLMs used in this study were among the most popular at the time. The evaluation process was facilitated by Big-AGI (https://big-agi.com/), an AI platform designed to make advanced artificial intelligence accessible. Big-AGI was chosen for its ease of integrating multiple models via API, the ability to implement system prompts, and its 'beam' function, which allows simultaneous submission of the same prompt to multiple LLMs.

The LLMs included in the study are as follows and in the version available in Italy in September 2024:
\begin{enumerate}
    \item GPT-4o: released in May 2024, GPT-4o is a multilingual, multimodal generative pretrained
transformer developed by OpenAI. 
Link: https://openai.com/\newline index/
hello-gpt-4o/
    \item Gemini 1.5 Pro Latest: a large language model developed by DeepMind (Google), is natively
multimodal and supports an extended context window of up to two million tokens, which is
currently the longest of any large-scale foundation model. Link: https://deepmind.google/technologies/gemini/pro/.
    \item Claude 3.5 Sonnet: developed by Anthropic, excels in the ability to understand nuanced language, humour, and complex instructions. 
Link: https://
www.anthropic.com/news/claude-3-5-sonnet.
    \item Mistral Large (2402): is designed to excel in complex reasoning tasks, particularly in multilingual contexts. The model is highly effective in text understanding, transformation, and code generation.
Link: https://mistral.ai/news/
mistral-large/.
    \item Open Mixtral 8x22B (2404): is one of the latest model developed by Mistral, featuring a sparse Mixture-of-Experts (SMoE) architecture. Despite its large size, with 141 billion parameters, only 39 billion parameters are actively engaged during processing, optimising both performance and cost efficiency. This approach sets new standards in the AI community for balancing model complexity with computational resource usage. Link: https://mistral.ai/news/
    mixtral-8x22b/.
    \item Llama 3.1 70B Instruct Turbo: developed by Meta, is a 70-billion parameter language model designed for instruction-following tasks. 
Link: https://
ai.meta.com/blog/meta-llama-3-1/
    \item Qwen2 72B Instruct: developed by Alibaba Cloud, is a 72-billion parameter language model
optimised for instruction-based tasks. Link:https://
www.alibabacloud.com/en/solutions/generative-ai/qwen?\_p\_lc=1 \newline (UNESCO, Webb, 2023).
\end{enumerate}

\paragraph{Prompting}
The aim of this study was to understand what models could be used by university teachers (and possibly other teachers or students). For this reason, overly sophisticated prompting techniques were not used, but rather what an educator might do by providing clear instructions and the necessary contextual data for evaluation.
The LLMs systems were prompted with the following instructions (originally written in Italian). The first is the system prompt:

\begin{quote}
{\ttfamily
You are an experienced and impartial university lecturer. Your job is to assess the quality of student assignments according to a specific assessment rubric. 

**How to respond to requests:**

* Do not express personal opinions or subjective judgements.
* Focus exclusively on the criteria provided in the rubric.
* Provide a fair and impartial assessment based on the task's \newline adherence to the criteria.
* Carefully review the student's \newline entire paper before beginning the assessment.
* Offer \newline constructive \newline suggestions as to how the student might improve.
* Uses clear and concise language.
* Justify the marks awarded with specific references to the paper and the rubric.
* In your assessment, take into account that the students only had 2 hours for planning.

**Request format

Each request will include:

* **The student's assignment:** The text of the assignment you are to assess.
* **The grading rubric:** A list of criteria with descriptions for each grade level.

**Response format:**

Your answer should follow this format:

**Title of the paper (also called title of the paper) as it \newline appears in the document: [insert title here]**.

**Total score:** [Insert total score here].

**Scoring breakdown:**

| Criterion | Score | Comments
|---|---|---|
| [Criterion 1] | [Score] | [Comments with specific examples from the task] |
| [Criterion 2] | [Score] | [Comments with specific examples from the task] |
| [Criterion 3] | [Score] | [Comments with specific examples from the task] |
| ... | ... | ... |

**Suggestions for improvement

* [Suggestion 1]
* [Suggestion 2]
* ...
**Answer following the answer format provided above.**
}
\end{quote}

The second is the  prompt that were given to the LLMs to assess the products (originally written in Italian):

\begin{quote}
{\ttfamily
Evaluate the attached teaching design (student task) that was created by a group of students from the secondary school teaching qualification course. The key competence of this assignment lay in being able to design a teaching intervention that makes effective use of teaching architectures and strategies. In particular, the group's competence in terms of redesign and depth of reflection is taken into account with respect to previous instructional design. At the same time, the instructional design had to prove effective in achieving the goals they set themselves. Take into account that the students only had 2 hours to design. Use the evaluation rubric below to assess:

<Starting teaching design evaluation rubric> 
[follows the rubric visible in Table~\ref{tab:assessment-rubric}]
<end of assessment rubric>

**Total score:** 

**Scoring distribution:**
}
\end{quote}

The prompt was sent simultaneously to all the LLMs involved. Through big-AGI the authentic task document was attached in PDF format. A zero-shot prompting procedure was used for all LLMs, meaning that no examples of human task assessments were given to the models.

\paragraph{Models’ Settings}

In order to set a common limit of length for every model’s answers, all of them have been set through big.AGI API controls to 8128 tokens maximum. Also, the temperature was set to 0.2, that should ensure  strict adherence to the instructions yet leave some room for creativity in answers.

\subsubsection{For Assessing LLMs' Feedback} 
We used a qualitative methodology for the analisys of the feedbacks from LLMs. We used as a basis the framework by Hughes, Smith, and Creese (2015), looking closely at the different parts of the feedback from the language models. We focused on praise, critique, advice, and questions. In February 2025, we tested three of the best LLMs to choose the best one to do the coding work. Once selected, we followed and adapted the Morgan (2023) protocol to keep our analysis clear and reliable.

\paragraph{Feedback form profiling instrument}
For the development of the content analysis, on the basis of the Hughes, Smith and Creese study (2015) a specific rubric was adopted in order to detect the presence of specific and crucial elements on the feedback produced by the LLMs selected. For their study, Hattie and Timperley's (2007) distinction between feedback and feedforward was integrated with Orsmond and Merry's (2011) feedback categories to create a new feedback profiling tool. This tool was designed to categorise feedback on a large scale within a social science context. The categories identified are: a) appreciation; b) recognition of progress; c) criticism; d) suggestion; e) questioning. The first step is strictly connected to the Hattie's and Timperley's last level and refers to positive comments that are usually intended to motivate students, although not very informative. The second category could be referred to the Hattie and Timperley area “appreciation”, but specifies more what kind of progress was made by the student, also pointing out that a lack of progress could be a problematic element.The third level allows to highlight criticism of the task in relation to the expected standards: due to the complexity of this category, the authors divided it in three sub-levels: c1) correction of errors, mostly referring to formal or grammatical errors; c2) criticism of the content, which refers to errors in the proposed arguments, and, finally, c3) criticism of the structural approach adopted in the task. The fourth level consists of suggestions for improvement, which may relate to: d1) specific elements relating to the current task, d2) more general elements still relating to the current task and d3) broader elements that can also be transferred to future work and learning activities. The final level, on the other hand, consists of asking the students for clarification on the less clear points and then taking action and engaging them in a dialogue to respond to these requests. 

\begin{table}[H]
\centering
\caption{Feedback profiling tool categories and examples reported from Hughes, Smith and Creese}
\label{tab:llms-assessment}
\renewcommand{\arraystretch}{1.2} 
\resizebox{\textwidth}{!}{ 
\begin{tabularx}{\textwidth}{|p{3cm}|p{6cm}|p{5cm}|}
\hline
\textbf{Element} & \textbf{Description} & \textbf{Example} \\
\hline
\multicolumn{3}{|l|}{\textbf{P. Praise} - Motivates students but may seem insincere if overused.} \\
\hline
P1. Praise & General positive feedback to encourage students. & "Excellent work on structuring your intervention clearly and logically. It demonstrates a solid understanding of the teaching objectives." (ID12, GPT-4o) \\
\hline
P2. Recognising Progress & Highlights student progress. Helps motivate and inform about learning gaps. & "Compared to the previous design, you have improved the clarity of the learning objectives." (ID10, Qwen2 72B) \\
\hline
\multicolumn{3}{|l|}{\textbf{C. Critique} - Identifies gaps but should include suggestions to improve.} \\
\hline
C1. Correction of errors & Points out mistakes (numerical, verbal, spelling, etc.). & “Objective 7, ‘Improve the ability to collaborate and communicate in groups’, could be reworded to be more specific and measurable.” (ID10, Qwen2 72B) \\
\hline
C2. Factual critiques & Highlights inaccuracies or content issues. & “The implementation of the debate strategy could be better structured to ensure effective discussion among students.” (ID3, Mistral Large) \\
\hline
C3. Critique of approach & Comments on the structural or methodological approach. & “The design does not contain a section devoted to critical reflection on redesign with respect to a previous intervention.” (ID21, Gemini) \\
\hline
\multicolumn{3}{|l|}{\textbf{A. Advice} - Guides future actions to improve student learning.} \\
\hline
A1. Specific to current assignment & Suggestions directly related to the current work. & “To further improve the scanning of the intervention, it would be useful to better specify the roles of students and teacher at each stage.” (ID10, GPT-4o) \\
\hline
A2. General for current assignment & General advice for the current assignment. & “Considering the limited time available, it would be appropriate to simplify certain steps of the activity.” (ID7, Gemini) \\
\hline
A3. Future learning & Advice applicable to future tasks and learning. & “In the future, you could explore more formative assessment strategies to monitor students' in itinere progress.” (ID26, Claude 3.5 Sonnet) \\
\hline
\multicolumn{3}{|l|}{\textbf{Q. Clarification Requests} - Prompts deeper thinking through questioning.} \\
\hline
Q. Clarification Requests & Encourages reflection or justifications. & “How do you plan to integrate student feedback into the assessment phase?” (ID40, Gemini) \\
\hline
\end{tabularx}
}
\end{table}

In detail, the profiling tool we used is composed as Table~\ref{tab:llms-assessment} suggests.

The scoring system is based primarily on the assumption that the specific category is present or not in the feedback description. So a score of 1 point is given to each category when the relevant qualitative code is detected in the feedback.
In order to analyse the qualitative comments produced by the seven LLMs, the study of Morgan (2023) and the related procedure for qualitative analysis through LLMs was taken into account.
The author’s investigation of two datasets stemmed from a broad interest in using themes to interpret qualitative data, selecting two unpublished datasets that were previously analysed.
When comparing these previous analyses to ChatGPT’s results, a truly "naïve" evaluation wasn’t possible because the author had already studied the datasets. However, to minimize bias, the researchers relied on open-ended prompts rather than directly asking ChatGPT about "themes" in the data. Morgan (2023) also used follow-up questions to refine the AI-generated responses. This method successfully produced relevant thematic content without the author applying their prior knowledge.  
Because we started the analysis of qualitative data which were already produced by LLMs systems through bigAgi platform, we adopted Morgans’ strategy to scaffold the inquiry process, using instead of open-ended questions, a more detailed prompt structure based on the Hughes, Smith and Creese (2015) framework for analysing feedback features and qualitative coding.

For this operation, GPT-4o Latest, Claude Sonnet 3.5 New and Gemini Pro Experimental 02-05 were initially adopted; after a testing process, it was decided to select the results of GPT-4o Latest that proved to be more complete and effective in terms of structure and general development. The prompt introduced for guiding the selected LLMs is the following:

\begin{quote}
{\ttfamily
Evaluate the attached descriptive feedback, which was produced with respect to a teaching task that was created by groups of students on the secondary school teaching qualification course. The key competence of this task lay in being able to design a teaching intervention that made effective use of teaching \newline architectures and strategies. In particular, the groups’ \newline competence in terms of redesign and depth of reflection was \newline taken into \newline account with respect to previous instructional design. At the same time, the instructional design had to prove effective in achieving the goals they set themselves. The students only had 2 hours to design. The following evaluation rubric was used to produce the specific evaluation and descriptive feedback:

<Starting teaching design evaluation rubric> 
Evaluation rubric: \\

--Here the Evaluation rubric in Table 1 was inserted--\\

Consequently, the feedback produced have been collected and on this specific occasion I would ask you to assess their \newline composition in light of Hughes, Smith and Creese's [4] framework in relation to the presence of the following elements\\

--Table 2 was inserted here.--\\

Please count the exact number of feedback present in the document (identified with “ID” followed by a corresponding number). 
\newline Furthermore, could you please analyse how many times the elements of of Hughes, Smith and \newline Creese's framework occurred in the text? Could you please extract for me the right number of quotations related to the elements identified by the Hughes, Smith and Creese’s framework (2015)? Specifically, in how many feedback these \newline elements are present? Please highlight the distribution related to their presence in each feedback. In relation to this data, please create a descriptive table.
}
\end{quote}

\section{Results}

Table~\ref{tab:results} and Figure~\ref {fig:llmsxfeat} below presents the comparison process between the results produced by the different LLMs adopted. The total number of the feedback analysed is 35.
It is important to remember that it was decided to ask LLMs for generic feedback to simulate how teachers might request feedback.

\begin{figure}[H]
    \centering
    \includegraphics[width=0.95\linewidth]{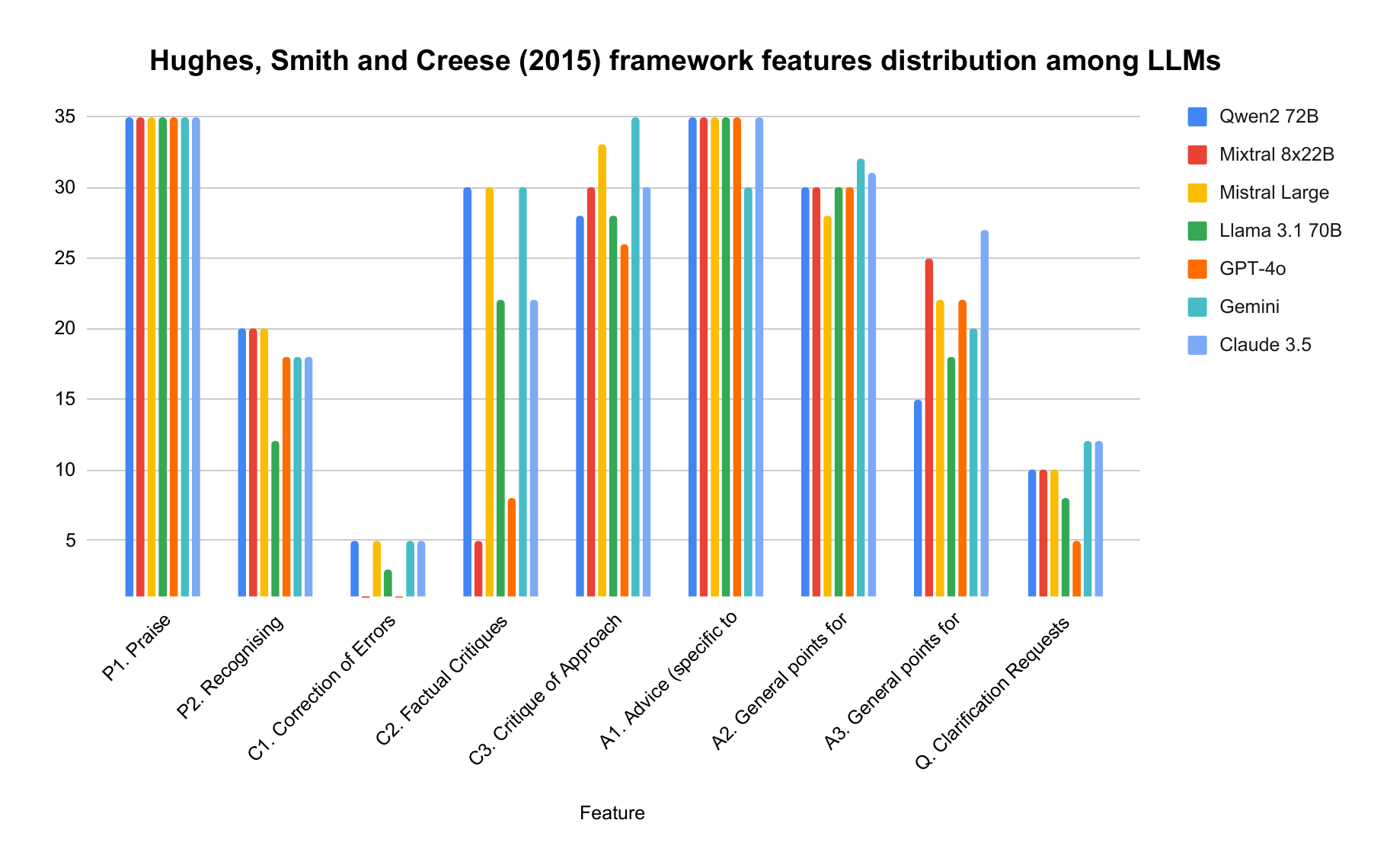}
    \renewcommand{\thefigure}{1}
    \caption{Hughes, Smith and Creese (2015) framework features distribution among LLMs}
    \label{fig:llmsxfeat}
    \renewcommand{\thefigure}{\arabic{figure}}
\end{figure}

\begin{table}[H]
\caption{Comparison of LLMs' feedbacks features. The maximum for each feature is 35, meaning that the relevant code has been detected in each one of the 35 feedbacks.}\label{tab:results}
\centering
\renewcommand{\arraystretch}{1.2}
\setlength{\tabcolsep}{4pt}
\begin{tabular}{|l|c|c|c|c|c|c|c|}
\hline
Element & Qwen2 & Mixtral & Mistral & Llama & GPT-4o & Gemini & Claude \\
\hline
P1. Praise (max 35) & 35 (100\%) & 35 (100\%) & 35 (100\%) & 35 (100\%) & 35 (100\%) & 35 (100\%) & 35 (100\%) \\
P2. Progress (max 35) & 20 (57\%) & 20 (57\%) & 20 (57\%) & 12 (34\%) & 18 (51\%) & 18 (51\%) & 18 (51\%) \\
C1. Errors (max 35) & 5 (14\%) & 0 (0\%) & 5 (14\%) & 3 (9\%) & 0 (0\%) & 5 (14\%) & 5 (14\%) \\
C2. Factual (max 35) & 30 (86\%) & 5 (14\%) & 30 (86\%) & 22 (63\%) & 8 (23\%) & 30 (86\%) & 22 (63\%) \\
C3. Approach (max 35) & 28 (80\%) & 30 (86\%) & 33 (94\%) & 28 (80\%) & 26 (74\%) & 35 (100\%) & 30 (86\%) \\
A1. Advice (max 35) & 35 (100\%) & 35 (100\%) & 35 (100\%) & 35 (100\%) & 35 (100\%) & 30 (86\%) & 35 (100\%) \\
A2. General (max 35) & 30 (86\%) & 30 (86\%) & 28 (80\%) & 30 (86\%) & 30 (86\%) & 32 (91\%) & 31 (89\%) \\
A3. Future (max 35) & 15 (43\%) & 25 (71\%) & 22 (63\%) & 18 (51\%) & 22 (63\%) & 20 (57\%) & 27 (77\%) \\
Q. Clarify (max 35) & 10 (29\%) & 10 (29\%) & 10 (29\%) & 8 (23\%) & 5 (14\%) & 12 (34\%) & 12 (34\%) \\
\hline
\end{tabular}
\end{table}

\begin{figure}
    \centering
    \includegraphics[width=1.2\linewidth]{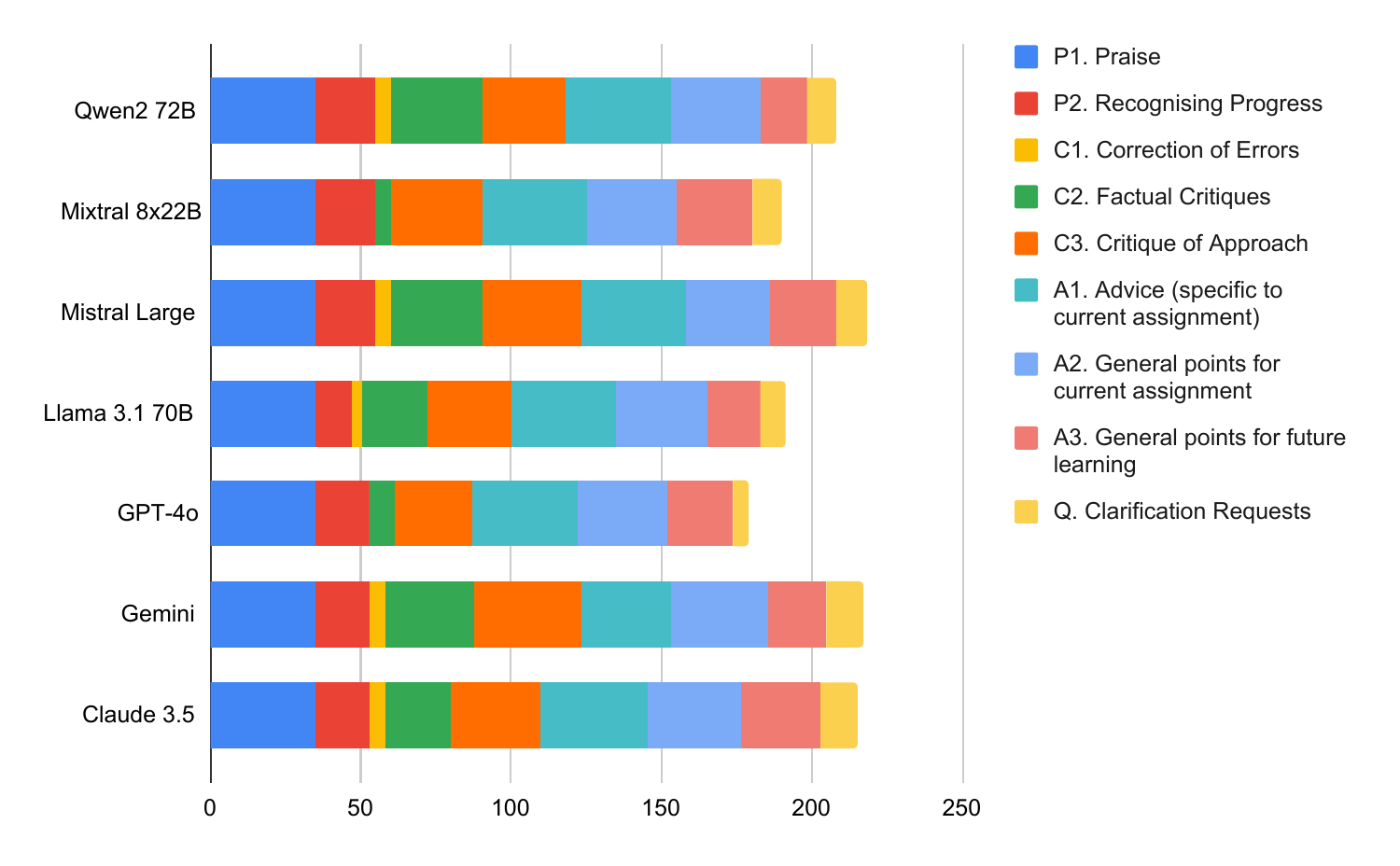}
    \caption{Performance score of LLMs' feedbacks based on the occurrences of feedback features}
    \label{fig:llmspoints}
\end{figure}

To conclude, the total scores related to the general areas of Hughes, Smith and Creese’ framework (2015) are summarised in Table~\ref{tab:summary} and Figure~\ref{fig:llmspoints}.

\begin{table}[H]
\caption{Summary of Total Scores and Accuracy for General Features and Overall}\label{tab:summary}
\centering
\renewcommand{\arraystretch}{1.2}
\setlength{\tabcolsep}{4pt}
\begin{tabular}{|l|c|c|c|c|c|c|c|}
\hline
Element & Qwen2 & Mixtral & Mistral & Llama & GPT-4o & Gemini & Claude \\
\hline
Total P (Praise, max 70) & 55 (78\%) & 55 (78\%) & 55 (78\%) & 47 (67\%) & 53 (76\%) & 53 (76\%) & 53 (76\%) \\
Total C (Critique, max 105) & 63 (60\%) & 35 (33\%) & 68 (65\%) & 53 (50\%) & 34 (32\%) & 70 (67\%) & 57 (54\%) \\
Total A (Advice, max 105) & 80 (76\%) & 90 (86\%) & 85 (81\%) & 83 (79\%) & 87 (83\%) & 82 (78\%) & 93 (88\%) \\
Total Q (Clarify, max 35) & 10 (28\%) & 10 (28\%) & 10 (28\%) & 8 (23\%) & 5 (14\%) & 12 (34\%) & 12 (34\%) \\
Total Score (Max 315) & 208 (66\%) & 190 (60\%) & 218 (69\%) & 191 (60\%) & 179 (57\%) & 217 (69\%) & 215 (68\%) \\
\hline
\end{tabular}
\end{table}

\section{Discussion}

\subsection {P Praise}

Looking at Figure~\ref{fig:llmsxfeat}, it is evident that all the LLMs successfully identified the general Praise (P1) element in each of the 35 pieces of feedback analysed. The framework authors (Hughes, Smith and Creese, 2015) highlight this element as “thought to be motivating for students, but if used indiscriminately it can appear insincere.” In line with this, Lipnevich and colleagues (2023) suggest that although praise can positively influence emotions, research has shown its dual effect: it can also diminish students' motivation and academic self-concept. These conflicting effects seem to occur mainly when teachers focus praise on personal traits or abilities rather than on effort and learning processes.

Regarding P2. Recognising Progress, which emphasises the importance of “acknowledging progress” to motivate and inform students about their learning, the LLMs that produced the most effective feedback were Qwen2 72B, Mixtral 8x22B, and Mistral Large. The analysis revealed examples such as:

\begin{quote} “Compared to the previous design, you have improved the clarity of the learning objectives (ID10)” (Qwen2 72B). \end{quote} \begin{quote} “The critical reflection is more thorough than the first design, but could be further improved (ID22)” (Mistral Large). \end{quote}

\subsection{C Critique}

Area C Critique focuses on identifying comments that help scaffold suggestions, recognising that “students need to know how their work falls short of expectations or criteria; however, criticism can be discouraging, especially when not accompanied by information on how to improve” (Hughes, Smith and Creese, 2015).

Regarding potential weaknesses in the LLMs, Correction of Errors (C1) was the least frequently observed element. Notably, Open Mixtral and GPT 4o did not address this aspect in any of their feedback. In contrast, Qwen2 72B (5/35) provided comments such as:

\begin{quote} “Objective 7, ‘Improve the ability to collaborate and communicate in groups’, could be reworded to be more specific and measurable (ID10).” \end{quote}

This was categorised as a correction of an error. It is noteworthy that the feedback was delivered politely and, more importantly, contextualised as an opportunity for improvement, thereby fostering inner-feedback and the development of critical thinking (Nicol, 2019).

Claude, in comparison, was more direct, offering comments like:

\begin{quote} “Beware of the use of the term ‘metacognitive-autoregulative’, which in this context might be more appropriate with a clearer definition (ID19).” \end{quote}

Mistral, however, was the most precise, as illustrated by:

\begin{quote} “In step 6, it might be useful to specify how students should reflect on their homework and how this reflection process will be evaluated (ID24).” \end{quote}

Regarding C2. Factual Critiques (of content), Qwen2 72B (30/35), Mistral Large (30/35), and Gemini (30/35) emerged as the most effective LLMs. For instance, Mistral Large offered highly specific feedback on the teaching methodology mentioned by students, introducing concise yet valuable formative suggestions:

\begin{quote} “The implementation of the debate strategy could be better structured to ensure effective discussion among students (ID3)” (Mistral Large). \end{quote}

Similarly, Gemini provided brief yet pointed remarks, such as:

\begin{quote} “The objective ‘to know the socio-historical context’ is too general and should be more specific (ID20).” \end{quote}

Qwen2 72B also delivered precise and direct feedback:

\begin{quote} “Learning objectives could be further aligned with the specific competences students should acquire (ID15).” \end{quote}

Critique of Approach (C3), which pertains to comments on the structural approach adopted in the task, was another well-covered element of the framework. All the LLMs addressed this aspect, with Gemini (35/35) and Mistral (33/35) demonstrating the highest effectiveness. For example, Gemini provided feedback such as:

\begin{quote} “The design does not contain a section devoted to critical reflection on redesign with respect to a previous intervention (ID21).” \end{quote}

Mistral Large also contributed valuable insights:

\begin{quote} “The implementation of the debate strategy could be better structured to ensure effective discussion among students (ID3).” \end{quote}

These examples illustrate the LLMs’ capability to provide precise and constructive feedback that effectively scaffolds students’ improvement.

\subsection{A Advice}

LLMs also demonstrate an effective capability to identify the area A Advice, which is defined as “important when the main purpose of feedback is to help students take future action to improve” [4]. The three domains within the Advice section—A1. Advice (specific to the current assignment), A2. General points for the current assignment, and A3. General points for future learning—are widely represented in the LLMs' feedback, potentially indicating a point of strength.

For A1., six out of seven LLMs (except for Gemini 30/35) included this element in all 35 pieces of feedback analysed. Some illustrative examples include:

\begin{quote} GPT 4o: “To further improve the scanning of the intervention, it would be useful to better specify the roles of students and teacher at each stage (ID10).” \end{quote} \begin{quote} Llama 3.1 70B Instruct Turbo: “Consider the integration of digital technologies to support learning and enhance the student experience.” \end{quote} \begin{quote} Mistral Large: “It is recommended to make the learning objectives more specific and measurable to improve alignment with the intervention (ID6).” \end{quote}

Regarding A2., the LLMs generally performed well, but Gemini (32/35) and Claude 3.5 (31/35) were notably more effective at providing advice to enhance the current assignment. Qwen2 72B, Open Mixtral 8x22B, Llama 3.1 70B, and GPT-4o performed similarly (30/35).

For instance, Gemini provided the following comment:

\begin{quote} “Considering the limited time available, it would be appropriate to simplify certain steps of the activity (ID7).” \end{quote}

Similarly, Claude 3.5 Sonnet suggested:

\begin{quote} “Consider integrating digital tools to support collaborative learning and metacognitive reflection (ID14).” \end{quote}

Both comments are concise yet provide specific prompts for group reflection on the development of their assignment.

Concluding the overview of area A Advice, the A3 element, which involves suggestions for future work and learning, was best addressed by Open Mixtral 8x22B (25/35) and Claude 3.5 Sonnet (27/35), closely followed by Mistral Large and GPT 4.o (22/35). Open Mixtral included Suggestions for Future Work (A3) in 71

\begin{quote} “Evaluators could provide more guidance for long-term improvement.” \end{quote}

Meanwhile, Claude 3.5 Sonnet offered forward-looking advice like:

\begin{quote} “In the future, you could explore more formative assessment strategies to monitor students' in itinere progress (ID26).” \end{quote}

\subsection{Q Clarification Requests}

Clarification Requests (Q) are generally less prevalent and harder to identify within the written feedback analysed. This area is minimally represented and pertains to “asking learners to think more deeply about their work and generate actions themselves, which can be achieved through questioning and dialogue” (Hughes, Smith and Creese, 2015).

Despite this, Gemini and Claude 3.5 outperformed the other models in this category (both 12/35). Examples of their comments include:

\begin{quote} “How do you plan to integrate student feedback into the assessment phase?” (ID40) (Gemini). \end{quote} \begin{quote} “How do you plan to handle any difficulties of students less used to collaborative work?” (ID33) (Claude 3.5 Sonnet). \end{quote}

These questions effectively stimulate deeper reflection and self-generated action from students, aligning with the intended purpose of Clarification Requests.

\subsection{General LLMs Overview Performance}

To provide a comprehensive understanding, an overview of LLMs' performances and results is essential (Figure~\ref{fig:llmspoints}). Assuming that every LLM applied Praise (P1) in every feedback analysed, the following general reflections are presented.

Qwen2 72B demonstrates balanced coverage across the framework elements: Advice (A1-35/35), Advice on the current assignment (A2-30/35), content criticism (C2-30/35), and structure criticism (C3-28/35). It is also one of the top three models in detecting C1. Correction of Errors (5/35). However, it shows limited requests for clarification (Q-10/35).

Open Mixtral 8x22B excels in P2. Recognising Progress (20/35) compared to other LLMs and performs well in A1. Advice (35/35) and A3. General points for future learning (25/35). Conversely, it shows weaker coverage in content criticism (C2-5/35) and error correction (C1-0/35), potentially limiting feedback effectiveness.

Mistral Large displays robust coverage in content criticism (C2-30/35) and structure criticism (C3-33/35), complemented by strong performance in A3. It ranks among the best LLMs in five out of nine framework elements, particularly excelling in P2. Recognising Progress (20/35), C1. Correction of Errors (5/35), and A1. Advice (35/35).

Meta Llama 3.1 70B is proficient in A1. Advice (35/35) but shows weaker coverage for recognising progress (P2-12/35) and requests for clarification (Q-8/35). Nonetheless, it maintains a balanced distribution across other elements.

GPT 4.0 exhibits strong coverage of A1. Advice (35/35) but weaker performance in error correction (C1-0/35) and content criticism (C2-8/35). However, it excels in A2. General points for the current assignment (30/35) and A3. General points for future learning (22/35) compared to other LLMs.

Gemini offers a well-balanced distribution, excelling in the C Critique area (C1-5/35; C2-30/35; C3-35/35), particularly in C3. Critique of approach (structure and argument) and A2. General points for the current assignment (32/35). It also performs well in Q. Clarification Requests (12/35), but could improve in recognising progress (P2-18/35) and A1. Advice (30/35).

Claude 3.5 Sonnet demonstrates comprehensive coverage, excelling in A1. Advice (35/35), structural criticism (C3-30/35), and A2. General points for the current assignment (31/35). It also leads in A3. Suggestions for future learning (27/35) and Q. Clarification Requests (Q-12/35), offering highly interactive and formative feedback alongside Gemini.

\subsubsection{Detailed Analysis of Mistral Large (2402) Feedback}

Cross-referencing feedback from the seven LLMs with the data in Tables 3 and 4, Mistral Large (2402) emerges as the most balanced and comprehensive model, effectively meeting Hughes, Smith, and Creese's (2015) criteria:

\begin{itemize}
    \item \textbf{High Presence of Constructive Criticism:} 
    \begin{itemize}
        \item C2. Criticism on content: 86\% of feedback, higher than most other models.
        \item C3. Criticism on structure and argumentation: 94\% of feedback, the highest among the models.
    \end{itemize}
    
    \item \textbf{Balanced Praise, Suggestions, and Criticism:} 
    \begin{itemize}
        \item P1. Praise: Present in 100\% of feedback, ensuring a motivating approach.
        \item A1. Specific suggestions for the current task: Present in 100\% of feedback, offering practical and actionable guidance.
    \end{itemize}
    
    \item \textbf{Focus on Future Suggestions:} 
    \begin{itemize}
        \item A3. Suggestions for future learning: Present in 63\% of feedback, higher than other models such as GPT-4.0 (51\%) or Llama 3.1 (18\%).
    \end{itemize}
    
    \item \textbf{Requests for Clarification:} 
    \begin{itemize}
        \item Q. Requests for clarification or further learning: Present in 29\% of feedback, relatively high compared to other models (e.g., GPT-4.0 with 14\%).
    \end{itemize}
\end{itemize}

\subsubsection{Examples of Mistral Large (2402) Feedback Comments}

\begin{itemize}
    \item \textbf{Critique on Structure and Argumentation (C3):} 
    \begin{quote}
        “Critical reflection on redesign is present, but could be deepened by providing more detail on the reasons for educational choices and possible difficulties and solutions.” (ID3)
    \end{quote}
    \begin{quote}
        “The scanning of the intervention is clear and well-structured, but some details are missing on the timing of each activity and how the transition between one phase and the next will be managed.” (ID4)
    \end{quote}

    \item \textbf{Content Critique (C2):} 
    \begin{quote}
        “Some learning objectives could be made more specific and measurable, for example by specifying geometric concepts that students should master.” (ID8)
    \end{quote}

    \item \textbf{Specific Suggestions for the Current Task (A1):} 
    \begin{quote}
        “Provide more specific examples of how teaching strategies will be implemented in the different phases of the intervention.” (ID2)
    \end{quote}

    \item \textbf{Suggestions for the Future (A3):} 
    \begin{quote}
        “To improve future interventions, it might be useful to integrate active learning strategies, such as problem-based learning.” (ID41)
    \end{quote}

    \item \textbf{Request for Clarification (Q):} 
    \begin{quote}
        “How do you plan to evaluate the effectiveness of the implementation of the selected teaching strategies?” (ID39)
    \end{quote}
\end{itemize}

\subsubsection{Comparison and Implications}

Mistral Large (2402) provides the most effective, balanced, and constructive feedback, with a focus on in-depth critique, practical suggestions, and future improvements. It offers a comprehensive mix of praise, criticism, and advice, making it highly beneficial for enhancing student learning. 

In comparison:
\begin{itemize}
    \item Claude 3.5 Sonnet provides balanced criticism, suggestions, and clarification requests, offering highly interactive and formative feedback.
    \item Gemini excels in quality criticism and practical suggestions.
\end{itemize}

Overall, these LLMs present significant opportunities for enhancing feedback processes. To maximise the benefits, it is crucial for university educators to effectively prompt AI to improve feedback quality (Estrada-Araoz et al., 2024). Thus, up-skilling processes are necessary to enhance teaching practices and fully leverage AI-powered learning (Pandeya and Kumaria, 2024). 

UNESCO (2023) highlights the importance of effective prompting:
\begin{itemize}
    \item Use clear and straightforward language.
    \item Include examples to illustrate desired responses.
    \item Provide context for generating relevant outputs.
    \item Refine and iterate prompts.
    \item Ensure ethical use by avoiding inappropriate or biased content.
\end{itemize}

Structured rubrics and criteria are also vital for guiding LLMs effectively, acknowledging that “AI can assist in the automated assessment of assignments and exams, providing instant feedback to students” (Owan et al., 2023; Estrada-Araoz et al., 2024).

\section{Conclusions}
This study has provided valuable insights into the potential and challenges of employing Large Language Models (LLMs) to generate educational feedback in higher education. Our analysis, grounded in Hughes, Smith, and Creese’s feedback framework (2015), reveals that while LLMs reliably produce well-formed praise and targeted advice, significant variations exist in their capacity to deliver nuanced corrective feedback and error detection. Notably, models such as Mistral Large demonstrated a balanced mix of content and structural criticism, underscoring the importance of selecting and calibrating AI tools based on specific pedagogical needs.
The implications of these findings are twofold. First, the integration of LLM-generated feedback could substantially alleviate the workload of educators, allowing for more timely and consistent formative assessments—a benefit aligned with the principles outlined by Hattie and Timperley (2015). Second, the observed disparities among models emphasise the need for careful rubric design and explicit prompting strategies to ensure that AI systems complement, rather than replace, the expert judgment of human instructors. This is critical for maintaining the quality and fairness of assessment processes, especially in light of the ethical considerations highlighted by UNESCO guidelines (2023).

To answer the research questions of this study, based on the analysis of the feedbacks, the study shows (RQ1) that LLMs can generate descriptive feedback that helps students learn. They produce responses that include praise, constructive criticism, and practical advice. This feedback guides students to recognize their strengths and identify areas for improvement. However, the effectiveness depends on clear instructions and context provided to the LLMs, and there is some variation between different models.
Additionally (RQ2), we can affirm that the feedback from the LLMs is well-formed. The analysis based on Hughes, Smith, and Creese’s framework (2015) demonstrates that the LLMs consistently include key elements—praise, critique, advice, and clarification requests—in their feedback. Although some models vary in specific aspects (for example, not all models address error correction equally), the overall structure and composition of the feedback meet the criteria for well-formed educational responses.

Despite its promise, the study also acknowledges several limitations. Variability in feedback quality, particularly in areas such as error correction and clarification requests, indicates that LLMs still require significant refinement before they can be fully integrated into high-stakes educational environments. Future research should explore methods to enhance the consistency of AI-generated feedback, including improved prompt engineering, adaptive rubric design, and longitudinal studies that assess the impact of such feedback on student learning outcomes (Agostini \& Picasso, 2024; Agostini, Picasso \& Ballardini,  2024). Additionally, it is essential to investigate how up-skilling initiatives for educators can bridge the gap between emerging AI technologies and traditional teaching practices (Estrada-Araoz et al., 2024; Pandeya and Kumaria, 2024).
Overall, while LLMs offer a promising avenue for enriching feedback processes in higher education, their effective deployment hinges on a synergistic relationship between technological innovation and pedagogical expertise. By addressing current limitations (that newer so-called ‘reasonong’ models, based chain-of-thoughts and test-time compute, might have already improved) and leveraging robust rubric frameworks or very specific prompting, stakeholders can harness the full potential of AI to foster meaningful learning, enhance academic performance, and support continuous pedagogical advancement (Panadero and Lipnevich, 2022; Hughes, Smith and Creese, 2025).

%
%

\begin{credits}

\subsubsection{\discintname}
Authors have no competing interests.
\end{credits}
%
%
%
%





\textbf{
\section{Disclosure statement} }
The authors report there are no competing interests to declare. 

\textbf{
\section{Data availability statement}
} 
The data that support the findings of this study are available from the corresponding author, [author's name] upon reasonable request. 

\vspace{1em}
\section*{References}

\begin{enumerate}[noitemsep]
    \item AGID Agenzia per l’Italia Digitale (2025). https://www.agid.gov.it/it/notizie/\newline intelligenza-artificiale-in-consultazione-le-linee-guida-pa.
    \item Agostini, D., \& Picasso, F. (2024). Large language models for sustainable assessment and feedback in higher education: Towards a pedagogical and technological framework. \textit{Intelligenza Artificiale}, \textit{18}(1), 121-138.   
    \item Agostini, D., Picasso, F., \& Ballardini, H. (2024). Large Language Models for the Assessment of Students' Authentic Tasks: A Replication Study in Higher Education. In \textit{CEUR Workshop Proceedings} (Vol. 3879). 
    \item Boud, D., \& Molloy, E. (2013). Rethinking models of feedback for learning: the challenge of design. \textit{Assessment \& Evaluation in Higher Education}, 38(6), 698-712.
    \item Bozkurt, A., Xiao, J., Farrow, R., John, Chrissi Nerantzi, Moore, S., Dron, J., Stracke, C. M., Singh, L., Crompton, H., Apostolos Koutropoulos, Evgenii Terentev, Pazurek, A., Nichols, M., Sidorkin, A. M., Costello, E., Watson, S., Mulligan, D., Honeychurch, S., \& Hodges, C. B. (2024). The Manifesto for Teaching and Learning in a Time of Generative AI: A Critical Collective Stance to Better Navigate the Future. \textit{Open Praxis}, 16(4), 487–513. https://doi.org/10.55982/openpraxis.16.4.777.
    \item Cardona, M.A., Rodríguez, R.J., \& Ishmael, K. (2023). \textit{Artificial Intelligence and Future of Teaching and Learning: Insights and Recommendations}, U.S. Department of Education, Washington, DC.
    \item Dochy, F. J. R. C., Segers, M., \& Sluijsmans, D. (1999). The use of self-, peer and co-assessment in higher education: A review. Studies in Higher education, 24(3), 331-350.
    \item Estrada-Araoz, E. G., Manrique-Jaramillo, Y. V., Díaz-Pereira, V. H., Rucoba-Frisancho, J. M., Paredes-Valverde, Y., Quispe-Herrera, R., \& Quispe-Paredes, D. R. (2024). Assessment of the level of knowledge on artificial intelligence in a sample of university professors: A descriptive study. \textit{Data and Metadata}, 3, 285.
    \item Gibbs, G., Simpson, C., \& Macdonald, R. (2003). Improving student learning through changing assessment - a conceptual and practical framework. \textit{European Association for Research into Learning and Instruction Conference, Padova, Italy}.
    \item Grion, V., \& Serbati, A. (2019). \textit{Valutazione sostenibile e feedback nei contesti universitari. Prospettive emergenti, ricerche e pratiche} (pp. 1-158). PensaMultimedia.
    \item GTnum (2023, April). Intelligence artificielle et éducation: Apports de la recherche et enjeux pour les politiques publiques. \textit{Carnet Hypothèses ‘Éducation, numérique et recherche’}. https://edunumrech.hypotheses.org/8726.
    \item Hattie, J., \& Timperley, H. (2007). The power of feedback. \textit{Review of Educational Research}, 77(1), 81-112.
    \item Hattie, J., \& Yates, G. C. (2013). \textit{Visible learning and the science of how we learn}. Routledge.
    \item Hughes, G., Smith, H., \& Creese, B. (2015). Not seeing the wood for the trees: developing a feedback analysis tool to explore feed forward in modularised programmes. \textit{Assessment \& Evaluation in Higher Education}, 40(8), 1079-1094.
    \item Kamalov, F., Santandreu Calonge, D., \& Gurrib, I. (2023). New era of artificial intelligence in education: Towards a sustainable multifaceted revolution. \textit{Sustainability}, 15(16), 12451.
    \item Kochmar, E., Vu, D. D., Belfer, R., Gupta, V., Serban, I. V., \& Pineau, J. (2022). Automated data-driven generation of personalized pedagogical interventions in intelligent tutoring systems. \textit{International Journal of Artificial Intelligence in Education}, 32(2), 323-349.
    \item Lipnevich, A. A., Eßer, F. J., Park, M. J., \& Winstone, N. (2023). Anchored in praise? Potential manifestation of the anchoring bias in feedback reception. \textit{Assessment in Education: Principles, Policy \& Practice}, 30(1), 4-17.
    \item Lipnevich, A. A., \& Panadero, E. (2021). A review of feedback models and theories: Descriptions, definitions, and conclusions. In \textit{Frontiers in Education} (Vol. 6, p. 720195). Frontiers,  (2021, December).
    \item Minn, S. (2022). AI-assisted knowledge assessment techniques for adaptive learning environments. \textit{Computers and Education: Artificial Intelligence}, 3, 100050.
    \item Morgan, D. L. (2023). Exploring the use of artificial intelligence for qualitative data analysis: The case of ChatGPT. \textit{International Journal of Qualitative Methods}, 22, 16094069231211248
    \item Nicol, D. (2019). Reconceptualising feedback as an internal not an external process. \textit{Italian Journal of Educational Research}, 71-84.
    \item Nicol, D. (2010). Guiding principles of peer review: improving written feedback processes in mass higher education. \textit{Assessment and Evaluation in Higher Education}, 35(5), 501-517.
    \item Nicol, D. J., \& Macfarlane‐Dick, D. (2006). Formative assessment and self‐regulated learning: A model and seven principles of good feedback practice. \textit{Studies in Higher Education}, 31(2), 199-218.
    \item Orsmond, P., \& Merry, S. (2011). Feedback alignment: effective and ineffective links between tutors’ and students’ understanding of coursework feedback. \textit{Assessment \& Evaluation in Higher Education}, 36(2), 125-136.
    \item Owan V., Abang K., Idika D., Etta E., Bassey B. (2023). Exploring the potential of artificial intelligence tools in educational measurement and assessment. \textit{EURASIA J Math Sci Tech Ed}, 19(8), 2307. https://doi.org/10.29333/\newline ejmste/13428.
    \item Panadero, E., \& Lipnevich, A. A. (2022). A review of feedback models and typologies: Towards an integrative model of feedback elements. \textit{Educational Research Review}, 35, 100416. 
    \item Pandeya, K., \& Kumaria, M. (2024). Up-Skilling Professors for the AI Classroom: A Competency-Based Approach. \textit{Inspiring Soul}, 34.
    \item Picasso, F., Agostini, D., \& Serbati, A. (2025). Exploring the Potential of AI for Authentic Assessment in Education: Towards a New Model of Interaction. In \textit{The Emerald Handbook of Active Learning For Authentic Assessment} (pp. 249-269). Emerald Publishing Limited. 
    \item Russell Group. (2023). New principles on use of AI in education. https://\newline russellgroup.ac.uk/news/new-principles-on-use-of-ai-in-education/
    \item Sadler, D. R. (1989). Formative assessment and the design of instructional systems. \textit{Instructional Science}, 18(2), 119-144.
    \item Shute, V. J. (2008). Focus on formative feedback. \textit{Review of Educational Research}, 78(1), 153-189.
    \item UNESCO: Report “Guidance for generative AI in education and research.” https://www.unesco.org/en/articles/guidance-generative-ai-education-and-research (2023).
    \item UCL (2023). \textit{Using generative AI (GenAI) in learning and teaching. Teaching \& Learning.} https://www.ucl.ac.uk/teaching-learning/publications/2023/sep/\newline using-generative-ai-genai-learning-and-teaching (2023, September 12).
    \item Webb, M. (2023). \textit{A Generative AI Primer. National Centre for AI. }https://\newline nationalcentreforai.jiscinvolve.org/wp/2024/01/02/generative-ai-primer/

\end{enumerate}

\section*{Appendices}
\begin{table}[H]
  \caption*{Rubric for the Assessment of the Educational Intervention}
  \label{tab:assessment-rubric}
  \begin{tabular}{p{0.15\textwidth}p{0.2\textwidth}p{0.2\textwidth}p{0.2\textwidth}p{0.2\textwidth}}
    \toprule
    \textbf{Assessment Criteria} & \textbf{Insufficient Level (1 point)} & \textbf{Sufficient Level (2 points)} & \textbf{Good Level (3 points)} & \textbf{Excellent Level (4 points)}\\
    \midrule
    \textbf{Understanding and applying educational architectures} & 
    Demonstrates a limited understanding of educational architectures, with applications not always appropriate or consistent. & 
    Shows a basic understanding of educational architectures, applying them generally correctly but with some uncertainties. & 
    Applies educational architectures correctly, with a good understanding of their use in the specific context. & 
    Demonstrates a thorough understanding of educational architectures, applying them in an innovative and contextually relevant manner. Clearly justifies the choices made.\\
    \midrule
    \textbf{Selection and implementation of teaching and learning strategies} & 
    The teaching strategies chosen are limited or not always appropriate for the objectives of the intervention. & 
    Uses some relevant teaching strategies, but their implementation could be more targeted or diversified in relation to the intervention goals. & 
    Selects and implements appropriate teaching strategies with a good correlation to the intervention goals. & 
    Selects and implements highly effective and diversified teaching strategies, perfectly adapted to the objectives and context of the intervention.\\
   \midrule
    \textbf{Definition of the intended learning outcomes} & 
    The objectives are vague, not measurable or not aligned with the chosen teaching architectures and strategies. & 
    The objectives are present but could be more specific or better aligned with the teaching architectures and strategies. & 
    The objectives are well-defined and generally aligned with the chosen teaching architectures and strategies. & 
    The objectives are clear, specific, measurable and perfectly aligned with the chosen teaching architectures and strategies.\\
    \midrule
    \textbf{Detailed scanning of the intervention} & 
    The scan is incomplete, unclear, or lacks a logical progression of activities. & 
    The scan is present but could be more detailed or better structured in some parts. & 
    The scan is clear and generally well-structured, with a good progression of activities. & 
    The scan is detailed, logical and well-structured, with a clear progression of activities and realistic timeframes.\\
    \midrule
    \textbf{Critical reflection on the redesign process} & 
    There is a lack of critical reflection on the changes made or the justifications are superficial. & 
    Includes some reflection on the changes, but the analysis could be more thorough. & 
    Provides good reflection on the changes, with clear links to the learning objectives. & 
    Provides deep and critical reflection on the changes made, clearly justifying each choice in relation to the learning objectives.\\
    \bottomrule
  \end{tabular}
\end{table}
\end{document}